\documentstyle[sprocl,epsf]{article}
\def\Journal#1#2#3#4{{#1} {\bf #2}, #3 (#4)}

\def\NPB{{\em Nucl. Phys.} B}

\def\PRD{{\em Phys. Rev.} D}

\def\be{\begin{equation}}
\def\ee{\end{equation}}
\def\bea{\begin{eqnarray}}
\def\eea{\end{eqnarray}}

\newcommand{\eq}[1]{(\ref{#1})}
\newcommand{\nn}{\nonumber}
\newcommand{\fr}{\frac}

\newcommand{\mg}{m_{H^{\pm}}^2}
\newcommand{\ma}{m_{A^0}^2}

\newcommand{\lt}{\left}
\newcommand{\rt}{\right}

\begin{document}
\title{
ENHANCEMENT OF LOOP-INDUCED $H^\pm W^\mp Z^0$ VERTEX}
\author{SHINYA KANEMURA}
\address{Theory Group, KEK\\
Tsukuba Ibaraki 305, Japan}
%
%

\maketitle
\abstracts{We discuss 
the non-decoupling effects on the loop-induced $H^\pm W^\mp Z^0$ vertex 
in the general two Higgs-doublet model.  
The decay process $H^\pm \rightarrow W^\pm Z^0$ is analyzed at one-loop level 
and possible enhancement of the decay width is explored. 
We find that a novel enhancement can be realized 
by the Higgs non-decoupling effects 
with large mass difference between charged and CP-odd Higgs bosons.     
The branching ratio of the process can be a few \% $\sim$ 10\%.     
Therefore the decay mode 
may be detectable at LHC or future $e^+e^-$ linear colliders (LC's).}

\section{Introduction}
The Higgs sector remains unknown. 
Although the minimal model (with one Higgs doublet) does not 
contradict any current experimental data, various purely 
theoretical motivations often expect the extended Higgs sectors 
\cite{hhg}. 
The charged Higgs boson $H^\pm$ and CP-odd Higgs boson $A^0$ 
are always introduced in such the extension.
Thus the detection of $H^\pm$ ( or $A^0$ ) 
is very important to confirm the extended Higgs sectors.
If $H^\pm$ is light, it may be detectable through 
the mode $H^\pm \rightarrow \tau \nu$ and $cs$. 
However, heavy $H^\pm$ enough to open $tb$ as a main decay mode 
may be elusive because of large QCD background.   
In such the case, we have to investigate the possibility of 
alternative decay modes to prove $H^\pm$ 
with the branching ratio enough to yield substantial events. 
The possible modes may be 
$H^\pm \rightarrow \tau\nu$, $h^0 W^\pm$,  
$W^\pm Z^0$ and $W^\pm \gamma$. 
Unfortunately, it has been known that the latter two modes 
disappear at tree level in general Higgs models with multi-doublet 
structures, in which they can be induced only at loop levels \cite{mepocap}. 
Thus these modes have been considered as a clear signal 
for {\it exotic} Higgs sectors (for example, including triplets).

In this talk, we discuss the loop-induced $H^\pm W^\mp Z^0$ vertex 
in the two Higgs-doublet model (2HDM) and MSSM \cite{kanemu}. 
We examine the possible enhancement of the vertex 
at one-loop level. 
The decay width of $H^\pm \rightarrow W^\pm Z^0$ 
is then calculated at one-loop level and 
the branching ratio is estimated in order  to see 
whether it can be enhanced enough to be detected at LHC or LC's. 
We find that a novel enhancement can occur 
by the Higgs non-decoupling effects in 2HDM
in the constraint from the present data. 
The branching ratio can be a few \% $\sim$ 10\% 
for $\tan \beta > 1$ at $m_{H^\pm} = 300$ GeV.  
In MSSM, the branching ratio is less than $0.01$ \% for $\tan \beta > 1$. 
Therefore, $H^\pm \rightarrow W^\pm Z^0$ may be useful as a  
probe of Higgs sectors between MSSM, 2HDM and {\it exotic} Higgs models.  

\newpage
\section{Property of The Model}
We here consider the model with a softly-broken discrete symmetry 
under $\Phi_1 \rightarrow \Phi_1$, $\Phi_2 \rightarrow - \Phi_2$ 
\cite{kanemu}.
This Higgs potential is popular one and covers the MSSM Higgs potential  
as a special case \cite{hhg}. 
We assume the soft-breaking mass parameter $\mu_3^2$ to be real.
This model includes the five massive physical scalar bosons; 
namely, the charged ($H^\pm$), CP-odd neutral ($A^0$) and two CP-even neutral 
ones ($h^0$ and $H^0$).
Three Nambu-Goldstone 
bosons $w^\pm$ and  $z^0$ are absorbed into $W_L^\pm$ and $Z_L^0$. 

In 2HDM, whether the internal heavy Higgs bosons are decoupled 
or not is a model dependent problem. 
The masses of $H^\pm$, $A^0$ and $H^0$ are naively 
expressed like $M_i^2 =  \lambda_i v^2 + {\cal O} (\mu_3^2)$, where 
$\lambda_i$ represent the linear combinations of the 
quartic-coupling constants.
If large $M_i$ are realized by $\mu_3^2$ with keeping $\lambda_i$ 
to be small, the effects of heavy Higgs bosons then tend to be decoupled from 
low-energy observables as seen in MSSM.
Alternatively, if $M_i$ are large due to the growing $\lambda_i$ 
with a small $\mu_3^2$, the decoupling theorem no longer work and  
the non-decoupling effects of the masses appear \cite{kanemu,toh}.

The Higgs sector, in general, does not have the custodial 
$SU(2)_V$ symmetry in 2HDM.
In terms of $2 \times 2$ matrices 
${\cal M}_i = \left(i \tau_2 \Phi_i^{\ast}, \Phi_i \right)$, 
($i = 1, 2$), the Higgs sector includes the term  
$\sim \lambda' \{{\rm tr}({\cal M}_2 \tau_3 {\cal M}_1^{\dagger})\}^2$.  
Such the term is not invariant under the transformation  
${\cal M}_i \rightarrow g_L^\dagger {\cal M}_i g_R$      
($g_{L,R} \in SU(2)_{L,R}$).  
Namely, the term breaks $SU(2)_R$ and thus the custodial $SU(2)_V$ symmetry 
explicitly.           
The coupling constant of this term is expressed as 
$\lambda' = (\mg - \ma)/v^2$. 
The explicit breaking of $SU(2)_V$ in the Higgs sector 
is measured by the mass difference between $A^0$ and $H^\pm$ 
\cite{kanemu}.

\section{Loop-Induced $H^\pm W^\mp Z^0$ Vertex}

We observe the absence of the tree $H^\pm W^\mp Z^0$ coupling 
in 2HDM (and MSSM). 
The coupling is expected to be generated in the kinetic part of 
the Higgs sector:   
${\cal L}_{\rm THDM}^{kin} = 
   \lt(D_\mu \Phi\rt)^\dagger D^\mu \Phi +  
   \lt(D_\mu \Psi\rt)^\dagger D^\mu \Psi$,  
where $\Phi$ and $\Psi$ are the Higgs doublets in the basis of 
the gauge-eigenstates \cite{georgi};
\begin{eqnarray}
   \Phi = \left( \begin{array}{c}
                w^+                                             \\
                 \frac{1}{\sqrt{2}} ( \phi^0 + v + i z^0 )
                   \end{array} 
                   \right), 
   \Psi = \left( \begin{array}{c}
                H^+                                              \\
                 \frac{1}{\sqrt{2}} ( \psi^0  + i A^0 )
                   \end{array} 
                   \right). \label{gb}
\end{eqnarray}
$\phi^0$ and $\psi^0$ are the linear combinations of $h^0$ and $H^0$.   
Since $\Psi$ does not have any vacuum expectation value, 
we can see the absence of the tree $H^\pm W^\mp Z^0$ coupling. 
It can be induced through the mixing between 
$\Phi$ and $\Psi$ at loop-levels.

We next discuss the non-decoupling effects of heavy particles 
on the $H^\pm W^\mp Z^0$ vertex \cite{kanemu}.  
The effective Lagrangian is \cite{mepocap} 
\begin{eqnarray}
  {\cal L}_{\rm eff} 
  &=& f_{H^+ W^- Z^0} H^+ W^-_\mu Z^\mu + {\rm h.c.} \nn\\
  &+& g_{H^+ W^- Z^0} H^+ F^{\mu\nu}_Z F_{\mu\nu}^W + {\rm h.c.}\nn \\
  &+& h_{H^+ W^- Z^0} \;\;  i \epsilon_{\mu\nu\rho\sigma} 
      H^+ F^{\mu\nu}_Z F^{\rho\sigma}_W  + {\rm h.c.}. \label{ef}
\end{eqnarray}
The coefficient $f_{H^+ W^- Z^0}$ is mass-dimension 1 and the others 
are dimension $-1$.  
The contributions of the heavy particles of the masses $M_i$ to the 
coefficients can be estimated at one-loop level as 
\begin{eqnarray}&&  f_{H^+ W^- Z^0} 
       \sim g \times \fr{g}{\cos \theta_W} \times 
                    \fr{M_i^2}{v} ( \times \ln M_i)
       \sim \fr{m_W m_Z}{v^3} \times M^2_i (\ln M_i), \nonumber\\      
&&  g_{H^+ W^- Z^0}, \; h_{H^+ W^- Z^0} 
     \sim \fr{m_W m_Z}{v^3} \times \ln M_i.
\end{eqnarray}  
Therefore the naive power-counting shows that there may be 
substantial non-decoupling effects  
of the heavy Higgs bosons as well as heavy fermions 
on the one-loop induced $H^\pm W^\mp Z^0$ vertex. 
 
The vertex is, however, 
 strongly constrained by the custodial $SU(2)_V$ symmetry.   
Each term in the Lagrangian \eq{ef} comes from the following 
each operator;  
\begin{eqnarray}
{\rm tr}\lt[ \tau_3 (D_\mu {\cal M})^\dagger (D^\mu {\cal N}) \rt],
{\rm tr}\lt[ \tau_3 {\cal M}^\dagger {\cal N} 
                             F_Z^{\mu\nu} F_{\mu\nu}^W \rt]  
{\rm or} \; i \epsilon_{\mu\nu\rho\sigma}
     {\rm tr}\lt[ \tau_3 {\cal M}^\dagger {\cal N}
                     F_Z^{\mu\nu} F^{\rho\sigma}_W \rt], \label{op}
\end{eqnarray} 
where $2 \times 2$ matrices ${\cal M}$ and ${\cal N}$ are defined by 
${\cal M} = (i\tau_2 \Phi^\ast, \Phi)$ and 
${\cal N} = (i\tau_2 \Psi^\ast,\Psi)$. 
All the operators \eq{op} are not invariant under $SU(2)_R$ and thus 
$SU(2)_V$, the vertex is induced according to 
the $SU(2)_V$ breaking in the model.
\section{Enhancement of Decay Process $H^\pm \rightarrow W^\pm Z^0$}
We proceed to examine the decay process $H^\pm \rightarrow W^\pm Z^0$.
The decay width in 2HDM (with Type II Yukawa coupling \cite{hhg}) 
with large $\Delta m ( = m_{A^0} - m_{H^\pm})$   
and also that in MSSM with heavy sparticles are shown in Fig 1. 
In MSSM, in addition to the Higgs decoupling property, 
$m_{A^0}$ is approximately degenerated with $m_{H^\pm}$. 
Thus the Higgs effects are small and 
the heavy fermion effects are dominant. 
Since the $H^\pm tb$ coupling constant consists of 
$m_t \cot \beta$ and $m_b \tan \beta$,  
the top-quark contributions are rapidly reduced 
for larger $\tan \beta$ \cite{mepocap}.  
On the other hand, in 2HDM with large $\Delta m$,  
a novel enhancement of the width is realized for large $\tan \beta$ 
because of the non-decoupling effects of Higgs bosons \cite{kanemu}. 

Let us consider the branching ratio of this decay mode in 2HDM next. 
The other decay modes included here are 
$H^\pm \rightarrow tb$, $\tau \nu$ and $h^0W^\pm$.
The parameters are fixed as $m_{h^0} = 140$ GeV, 
$m_{H^0} = 310$ GeV and $\alpha = \beta - \pi/2$. 
We also assume
\newpage
\hspace*{1cm}
\begin{minipage}[t]{8cm}  
\epsfxsize=8cm
\epsfbox{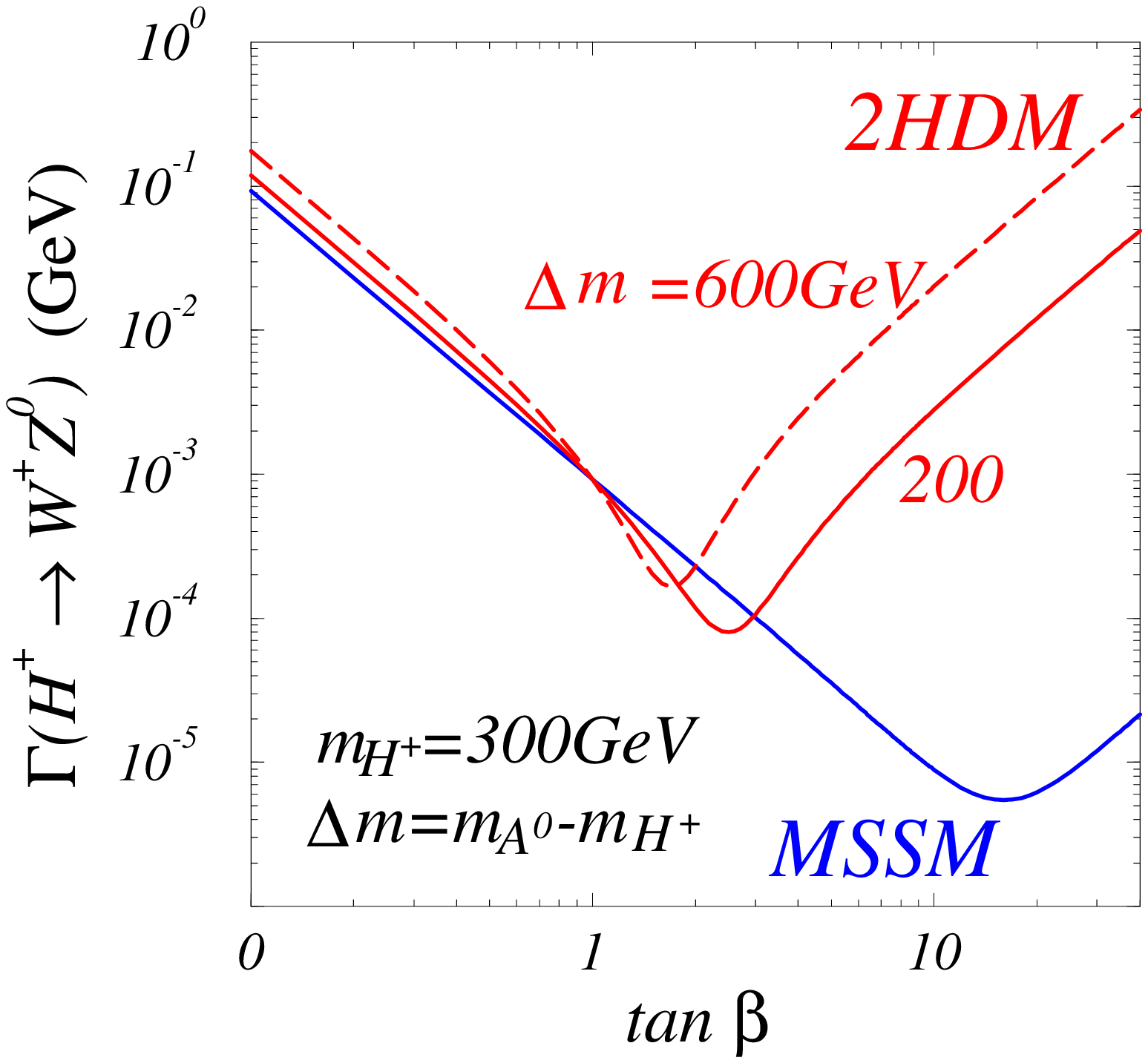}
\center{\bf Fig 1}
\end{minipage}\\ \\
\noindent
that $m_t = 175$ GeV and $m_b (m_{H^\pm}) = 3$ GeV.  
We can see in Fig 2 that the branching ratio become larger than $1$ \% 
if $\Delta m$ is greater than 200 GeV for $\tan \beta > 5 \sim 8$.  
The maximal value can amount to near $10$ \% for very large $m_{A^0}$ and 
$\tan \beta > 20$. 
In the nearly $SU(2)_V$ symmetric cases in the Higgs sector 
($m_{A^0} \sim m_{H^\pm} = 300$ GeV), 
the Higgs non-decoupling effects 
are canceled out and only the fermion effects remain,  
so that the branching ratio becomes less than $0.01$ \%. 
As mentioned before,
the non-zero $\mu_3^2$ reduces the non-decoupling Higgs effects. 
So far we have tried to extract the Higgs non-decoupling effects 
as large as possible, 
assuming the soft-breaking parameter $\mu_3^2$ to be zero.  
However, $\mu_3^2$ is often very important in various aspects of physics.   
The reduction of the branching ratio by 
$m_3 = \mu_3/\sqrt{\sin \beta \cos \beta}$ is shown in Fig 3. 
\section{Summary and Discussion}
We have discussed the loop induced $H^\pm W^\mp Z^0$ vertex. 
The possibility of its enhancement has been explored. 
The conditions for the enhancement is summarized as 
1) Non-decoupling properties of Higgs sector with small $\mu_3^2$,  and 
2) Large SU(2) breaking in the Higgs sector by the mass difference 
between $A^0$ and $H^\pm$. 
Although such conditions are not satisfied in the framework of MSSM, 
it is possible for 2HDM to satisfy them within the arrowed 
region from the current experimental data and also from the 
perturbative unitarity. 
As a result, the branching ratio of $H^\pm \rightarrow W^\pm Z^0$ 
can be a few \% $\sim$ 10 \%. 
Such the enhancement may make it possible to detect the decay mode 
at LHC and also \newpage
\noindent
\begin{minipage}[t]{8cm}  
\epsfxsize=8cm
\epsfbox{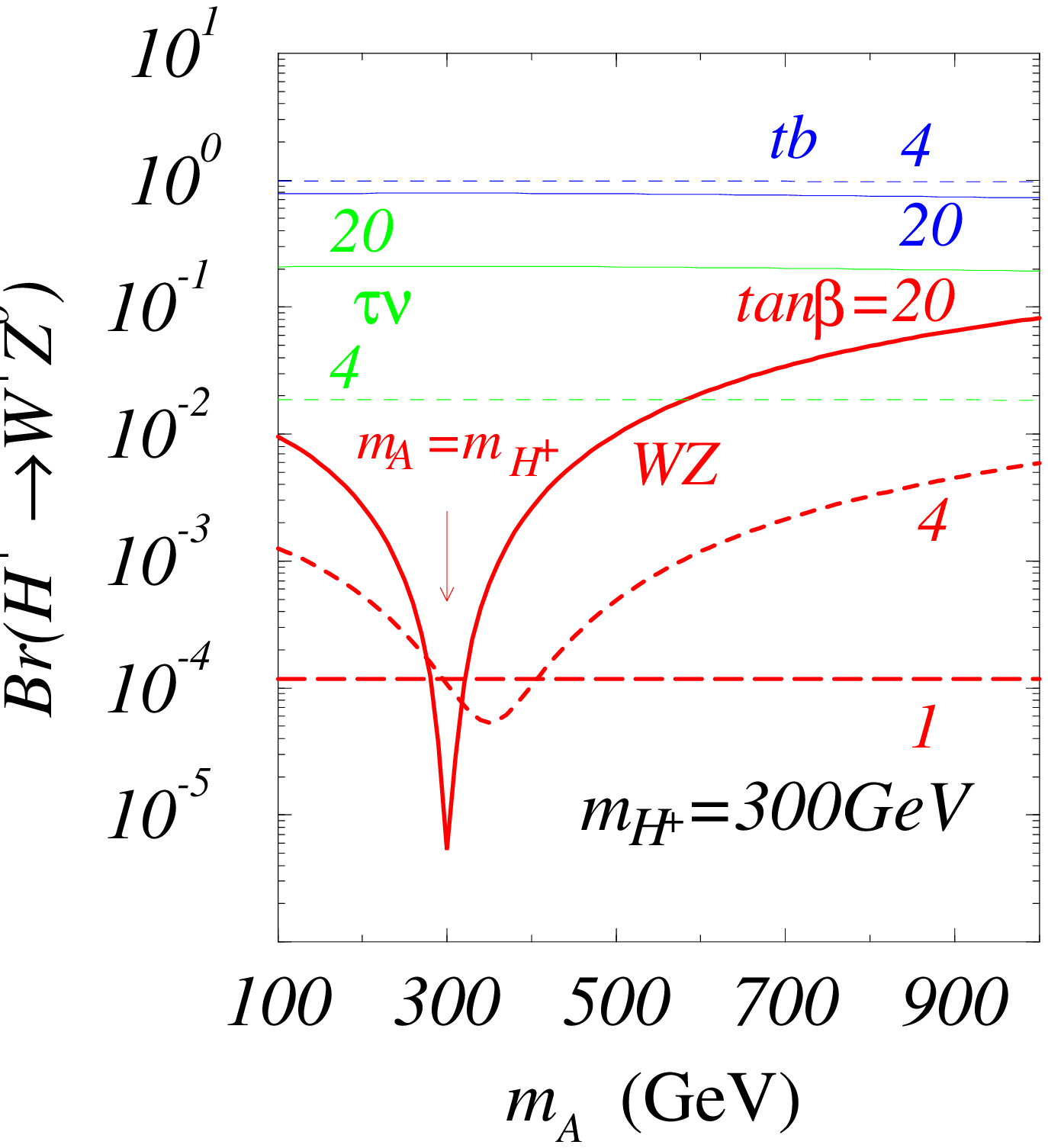}
\center
\hspace*{-1cm}
{\bf Fig 2}
\end{minipage}
\hspace{-2cm}
\begin{minipage}[t]{7.5cm}  
\epsfxsize=7.5cm
\epsfbox{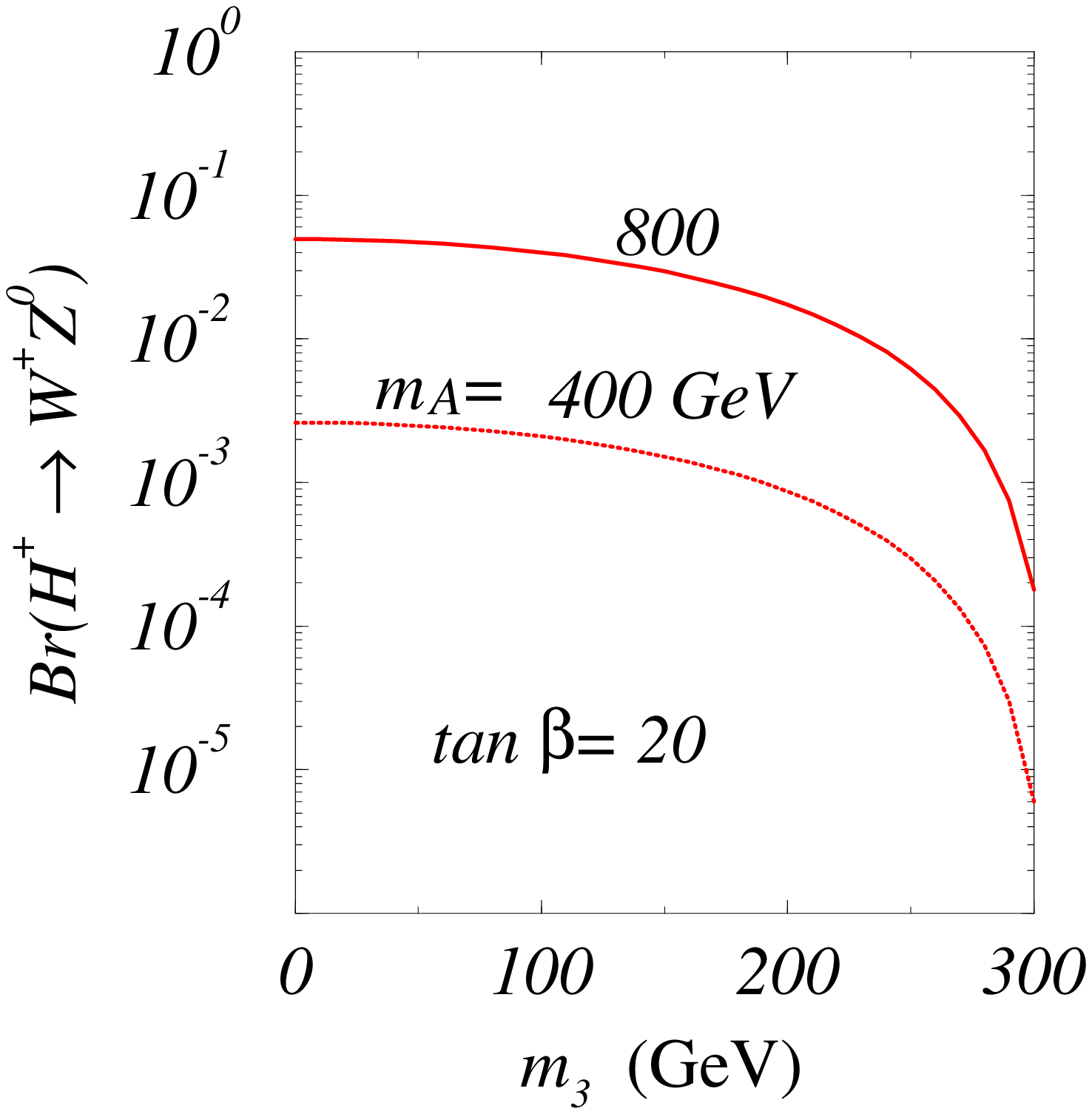}
\center{\bf Fig 3}
\end{minipage}\\ \\
\noindent
LC's. 
At LHC, the charged Higgs boson is mainly produced through the subprocess 
$g b \rightarrow t H^\pm$, where  over one hundred production of 
$H^\pm \rightarrow W^\pm Z^0 \rightarrow lll\nu$  
can be expected per a year for $Br(H^+ \rightarrow W^+ Z^0) > 1 $\%.  
Since the background (mainly $ud \rightarrow W Z$) 
is naively estimated to be such that  
a few \% of the branching ratio 
are required to see a signal, we can expect to detect 
the decay mode with the enhancement above.
We also expect that it can be detectable at LC's 
($\sqrt{s}= 1$ TeV, $L=160$ fb$^{-1}/$year), where over 
a few dozen events of 
$H^\pm \rightarrow W^\pm Z^0 \rightarrow lll\nu$, $lljj$, $l\nu jj$ 
and $\nu\nu jj$
are produced par a year for $Br(H^+ \rightarrow W^+ Z^0) > 1$ \% 
with less background.
\vspace*{-1mm}            
\section*{Acknowledgments}
\vspace*{-1mm}            
The author would like to A. Miyamoto for much valuable discussion.

\end{document}